\newif\ifpdf

\documentclass{aa}

\ifx\pdfoutput\undefined\pdffalse\else\pdfoutput=1\pdftrue\fi

\usepackage{amsmath,natbib,graphicx}

\ifpdf
  \usepackage{hyperref}
\fi

  \graphicspath{{pics/}}    

\newcommand{\B}[1]  {{\boldsymbol{#1}}}
\newcommand{\CR}    {{\mathrm{CR}}}
\newcommand{\olr}   {{\mathrm{OLR}}}
\newcommand{\ilr}   {{\mathrm{ILR}}}
\newcommand{\diff}  {{\mathrm{d}}}

\begin{document}
\title{Kinematic response of the outer stellar disk to a central bar}

\author{G. M\"uhlbauer$^1$ \and W. Dehnen$^{1,2}$} 
\authorrunning{G. M\"uhlbauer \and W. Dehnen} 
\offprints{W.~Dehnen}
\institute{
  $^1$ Max-Planck Institut f\"ur Astronomie, K\"onigstuhl 17,
        D-69117 Heidelberg \\
  $^2$ Astrophysikalisches Institut Potsdam, An der Sternwarte 16, 
        D14482 Potsdam \\
  \email{muehlbau@mpia.de, wdehnen@aip.de}
  }
\date{Received 10 December 2002 / Accepted 13 February 2003}

\abstract{ We study, using direct orbit integrations, the kinematic response of
  the outer stellar disk to the presence of a central bar, as in the Milky-Way.
  We find that the bar's outer Lindblad resonance (OLR) causes significant
  perturbations of the velocity moments. With increasing velocity dispersion,
  the radius of these perturbations is shifted outwards, beyond the nominal
  position of the OLR, but also the disk becomes less responsive. If we follow
  \citet{de00a} in assuming that the OLR occurs just inside the Solar circle and
  that the Sun lags the bar major axis by $\sim20^\circ$, we find (1) no
  significant radial motion of the local standard of rest (LSR), (2) a vertex
  deviation of $\sim10^\circ$ and (3) a lower ratio $\sigma^2_2/\sigma^2_1$ of
  the principal components of the velocity-dispersion tensor than for an
  unperturbed disk. All of these are actually consistent with the observations
  of the Solar-neighbourhood kinematics. Thus it seems that at least the
  lowest-order deviations of the local-disk kinematics from simple expectations
  based on axisymmetric equilibrium can be attributed entirely to the influence
  of the Galactic bar.   } \maketitle
\section{Introduction} \label{sec:intro}
Several observations since the 1980s have shown beyond any doubt that the Milky
Way is actually a barred galaxy. Although there is strong evidence that bars in
galaxies are confined to regions within the radius $R_\CR$ of the co-rotation
resonance (CR), a bar may nonetheless influence the outer parts of its host
galaxy, most obviously by resonant phenomena. It has been shown by \citet{de00a}
and \citet{fux01} that if the Sun is just outside the OLR such influence can
explain the so-called $u$-anomaly. The latter is a bi-modality in the local
velocity distribution of old stars, that has been inferred from HIPPARCOS data
\citep{de98a}.

The galactic bar might also have some relevance in explaining the vertex
deviation. This is the mis-alignment of the local velocity dispersion ellipsoid
with respect to the local radial direction in the Galaxy and vanishes for any
axisymmetric stellar dynamical equilibrium. Using HIPPARCOS data, the vertex
deviation has been shown \citep{db98} to reach as high as $30\degr$ for young
stellar populations, and to be $\sim 10\degr$ for the old ones. Whereas the high
vertex deviation of the young populations is most probably due to moving groups,
i.e.\ deviations from dynamical equilibrium, and thus in a sense accidental
\citep{BM}, there should be dynamical reasons for its occurrence with the old
populations, i.e.\ deviations from axisymmetry.

Another anomaly in the stellar kinematics observed in the Solar neighbourhood
that may also be related to the bar's influence is the low value for the ratio
$\sigma_2^2/\sigma_1^2$ of the principal components of the velocity-dispersion
tensor of only $0.42^{+0.06}_{-0.04}$ for old stellar populations \citep{db98}.
For a flat rotation curve for the Milky Way and in the limit of vanishing
velocity dispersion, epicycle theory gives a lower limit of 0.5 for this ratio
and the actual expectations from axisymmetric models are $\approx0.6$
\citep{ec93,de99a}, significantly inconsistent with the data.

In the present work, we will follow the approach of \citet{de00a} further by
investigating the velocity distribution in its spatial variability. Our primary
interest is in the low-order moments of this distribution, i.e.\ the mean
(streaming) velocity and velocity dispersion. In particular, we want to quantify
whether the influence of the Galactic bar may explain the aforementioned
anomalies, the vertex deviation and velocity-dispersion axis ratio, observed for
the old stellar populations. Our approach applies to the kinematics in the solar
neighbourhood, while at the same time it constitutes a completely general
analysis of the bar influence in a stellar disk.

The outline of the paper is as follows: In Sect.~\ref{sec:descrip} we will
describe our simulations, Sect.~\ref{sec:results} gives a rather technical
presentation of the results, and Sect.~\ref{sec:disc} concludes the paper by a
discussion of the results.

\section{Simulation of Bar Influence} \label{sec:descrip}
Since we do not want to construct a self-consistent model of the Galaxy, we just
study the stellar dynamics for a simple model potential. In order to arrive at
(stationary) equilibrium, we slowly add to the underlying axisymmetric Galactic
potential the non-axisymmetric component of the bar (the bar monopole is assumed
to be already accounted for by the Galactic potential). We do not pay very much
attention to the inner parts (inside co-rotation), and we neglect influences of
vertical motion, so our model is two-dimensional. The calculation consists of
orbit integration of a large ensemble of phase-space points representing the
initial equilibrium. This numerical technique, which may be called
\emph{restricted $N$-body method}, is equivalent to first-order perturbation
theory, since the self-gravity due to the wake induced by the perturbation (bar)
is neglected. After the orbit integrations, the velocity moments at certain
times are computed from the phase-space positions of the trajectories.
\subsection{Sampling} \label{sec:sampling}
The sampling of the initial phase-space points is done as described in
\citet{de99a}, using the distribution function \citep[eq. (10) in][]{de99a}.
\begin{equation} \label{eq:inifialDF}
  f(E,L) = \frac{\Omega(R_E)\,\Sigma(R_E)}{\pi\kappa(R_E)\,\sigma^2(R_E)}
        \exp\left[\frac{\Omega(R_E)[L-L_c(R_E)]}{\sigma^2(R_E)}\right],
\end{equation}
where $\Omega(R)$, $\kappa(R)$, and $L_c(R)$ are the azimuthal and epicycle
frequency and angular momentum of the circular orbit at radius $R$, while $R_E$
is the radius of the circular orbit with energy $E$. With this choice for the
distribution function, the collisionless Boltzmann equation is satisfied at
$t=0$ (since $f$ depends only on the integrals of motion $E$ and $L$) and the
surface density and radial velocity dispersion of the disk follow approximately
\citep[for a quantitative comparison, see][and also Fig.\ 
\ref{fig:zerocomps}]{de99a} those given with the \emph{parameter functions}
$\Sigma(R)$ and $\sigma(R)$, respectively. Here, we assume exponentials for both
of them:
\begin{equation}
  \label{sigmarun}
  \Sigma(R) = \Sigma_0\,{\rm e}^{(R_0-R)/R_\Sigma},\qquad
  \sigma(R) = \sigma_0\,{\rm e}^{(R_0-R)/R_\sigma}.
\end{equation}
If not stated otherwise, we choose $R_\sigma = R_0$ and $R_\Sigma = 0.33 R_0$ by
standard, $R_0$ being the distance of the Sun from the galactic center. If not
stated otherwise, $\sigma_0=0.2v_0$, where $v_0$ is the circular velocity at
$R_0$. In this way, samples of $K=10^7$ initial phase-space points are created.
\subsection{Model Potential and Orbit Integration} \label{sec:orbit}
\begin{figure}
  \centering
  \includegraphics[width=75mm]{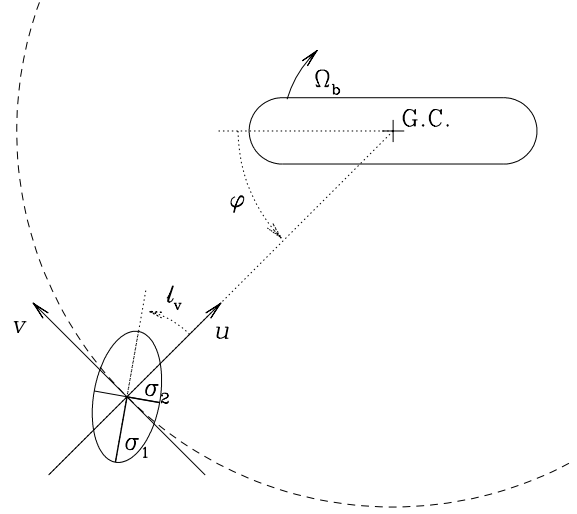}
  \caption{Geometry of the rotation of the galaxy and definition of coordinate
    system. Note that positive values of radial velocity u are taken to point
    inwards and that azimuth angle $\varphi$ is measured from the bar axis in
    the mathematically positive sense, but against the direction of bar rotation
    (modulo $180\degr$). Also shown is a velocity dispersion ellipsoid with its
    principal components $\sigma_1$ and $\sigma_2$ and a (positive) vertex
    deviation $\ell_v$.}
  \label{fig:coorddef}
\end{figure}
Orbit integration and adiabatic growth of a quadrupole bar is done similarly to
\citet{de00a}. The galactic background potential is chosen to give a power law
in the velocity curve
\begin{equation}
  v_c(R) = v_0  \left(R/R_0\right)^{\beta},
\end{equation}
namely:
\begin{equation} \label{galpot}
  \Phi_0(R) = \left\{
    \begin{array}{lr}
      \displaystyle
      (2\beta)^{-1} v_0^2 \left(R/R_0\right)^{2\beta}
      & \qquad {\rm for}\quad \beta \neq 0 \\ [1ex]
      \displaystyle
      v_0^2 \ln \left(R/R_0\right)
      & \qquad {\rm for}\quad \beta = 0 
    \end{array} \right.
\end{equation}
For the bar potential $\Phi_1 = \Phi_1(R, \varphi, t)$, we only use a
quadrupole, since higher poles are much less important at large radii, following
\citet{de00a}:
\begin{equation}
  \label{barpot}
  \Phi_1 = - A_b(t) \cos2\varphi\times\left\{
    \begin{array}{lr}
      \displaystyle
      2-\left( R/R_b \right)^3
      & \quad {\rm for}\; R \le R_b \\ [1ex]
      \displaystyle
      \left( R_b/R \right)^3
      & \quad {\rm for}\; R > R_b
    \end{array} \right.
\end{equation}
where $R_b$ is the size of the bar. The angle $\varphi$ is defined in the frame
rotating at pattern speed $\Omega_b$ (see Fig.\ \ref{fig:coorddef}). The strength
$A_b$ of the bar is increased from 0 to a value $A_f$ during time $0 < t < t_1$
according to
\begin{equation}
  A_b(t) = A_f \left(\frac{3}{16} \xi^5 - \frac{5}{8} \xi^3 + \frac{15}{16} \xi + 
    \frac{1}{2}\right), \quad \xi = 2 \frac{t}{t_1} - 1
\end{equation}
and stays constant at $A_b = A_f$ after $t_1$. With this functional form,
$\Phi_1$ and its first and second time derivative are continuous, thus
representing adiabatic growth of the bar and ensuring a smooth transition. The
final bar strength $A_f$ is controlled via the dimensionless model parameter
$\alpha_0=\alpha(R_0)$ with
\begin{equation} \label{alphadef}
  \alpha(R) = 3 \frac{A_f}{v_0^2} \left(\frac{R_b}{R}\right)^3,
\end{equation}
which is the ratio of the forces due to $\Phi_1$ and $\Phi_0$ at galacto-centric
radius $R>R_b$ on the bar's major axis. Our standard choice is $\alpha_0=0.01$,
which is consistent with the data for the Milky Way.

Orbit integration is performed in the co-rotating frame up to a time $t_2>t_1$
using a 5th order integrator. In contrast to \citet{de00a}, we integrate forward
in time.
\subsection{Calculation of Velocity Moments} \label{sec:moments}
The $(p,q)$th moment of the velocity distribution at position $\B{r}$ is defined
as
\begin{subequations} \label{eq:moments}
  \begin{eqnarray}
    \label{eq:moments-a}
    M_{pq}(\B{r},t)
    &=& \int\diff u\;\diff v\; u^p v^q \;f(t,\,\B{x}=\B{r},\,\B{v}) \\
    \label{eq:moments-b}
    &=& \int\diff\B{x}\;\diff\B{v}\;\delta(\B{r}-\B{x})\; u^p v^q\;
    f(t,\,\B{x},\,\B{v})
  \end{eqnarray}
\end{subequations}
with $\B{v}\equiv(u,v)$, where $u$ and $v$ denote, respectively,the radial and
azimuthal velocity component (see Fig.\ \ref{fig:coorddef}).  We do not know the
value $f(t,\,\B{x},\,\B{v})$ of the distribution function at any time $t>0$, but
instead have a representative sample $\{\B{x}_k(t), \B{v}_k(t)\}$ of phase-space
points. Hence, we compute the moment integral (\ref{eq:moments-b}) via
Monte-Carlo integration, resulting in
\begin{equation}\label{eq:moms}
  M_{pq}(\B{r},t) = \frac{M_{\mathrm{tot}}}{K}\,\sum_k  \frac{1}{\epsilon^2_k}\,
        w\left(\frac{|\B{x}_k - \B{r}|}{\epsilon_k}\right)\,u_k^p\, v_k^q.
\end{equation}
where we have replaced the $\delta$ function in Eq.~(\ref{eq:moments-b}) by
$\epsilon_k^{-2} w(|\B{x}_k-\B{r}|/\epsilon_k)$ with the weight function
\begin{equation}
  w(d) = \frac{2}{\pi} (1-d^2)\; \theta(1-d^2),
\end{equation} 
where $\theta$ is the Heaviside function. The parameter $\epsilon_k$ gives the
radius over which the $k$th trajectory contributes to the moment integrals. We
adjusted $\epsilon_k$ such that a constant number of sampled orbits was expected
to fall in the area of radius $\epsilon_k$ centered on $\B{x}_k$ by the assumed
exponential surface brightness distribution of the disk.

In practice, the zeroth, first and second moments are estimated in this way and
the mean velocities and velocity dispersion tensor\footnote{Notation:
  $\sigma^2_{uu}$ etc.\ are components of the tensor $\boldsymbol{\sigma}^2$,
  whose eigenvalues are the \emph{squares} of the velocity dispersions in the
  principal directions. Note that $\sigma^2_{uv}$ may well be negative.}  are
then evaluated from $\mu_{pq}\equiv M_{pq}/M_{00}$ as $\bar{u}=\mu_{10}$,
$\bar{v}=\mu_{01}$ and $\sigma^2_{uu}=\mu_{20}-\mu_{10}^2$,
$\sigma^2_{vv}=\mu_{02}-\mu_{01}^2$, $\sigma^2_{uv}=\mu_{11}-\mu_{10}\mu_{01}$.

Once the orbit integration is done, one can easily switch to a model with
initial distribution function $f_1\neq f$ by weighting each orbit with the ratio
$f_1/f$, which accounts for the fact that the trajectories were actually sampled
from $f$.  However, for this method to be useful, the ratio $f_1/f$ must not
become too large, because otherwise the moment estimates are dominated by a few
orbits with large weights. Since, by virtue of the collisionless Boltzmann
equation, the values of the distribution functions are conserved along
trajectories, the ratio $f_1/f$ is conserved, too, and can be evaluated at time
$t=0$, for which the distribution functions can be computed via
Eq.~(\ref{eq:inifialDF}). In this way, we did a switch to exponential disks with
velocity dispersions $\sigma_0<0.2v_0$ (the value of the sampling distribution
function), such that $f_1/f \le [0.2v_0/\sigma_0]^2$ (this follows from
Eq.~(\ref{eq:inifialDF}); for $\sigma_0> 0.2v_0$, the ratio $f_1/f$ can become
very large)
\begin{figure}
  \centering \includegraphics[width=80mm]{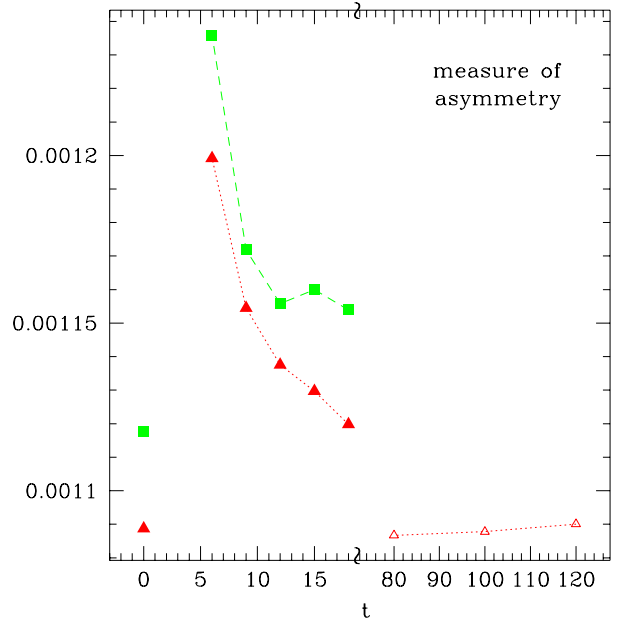}
  \caption{Evolution
    towards stationarity as measured by the symmetry (\ref{phisym}) for
    different models (arbitrary units, horizontal axis is time in bar rotation
    periods). Models shown have $R_\olr/R_0 = 0.8$ ({\em squares}) and
    $R_\olr/R_0 = 0.92$ ({\em triangles}), and usual ({\em filled}) or long
    integration time ({\em open}). Bar growth is taking place in the first 5
    periods for normal and in the first 10 periods for the long integration
    time. In addition, the value for the sampled starting distribution at $t=0$
    is shown, indicating the noise level.}
  \label{fig:symtest}
\end{figure}
\subsection{Fourier Components of Velocity Moments}\label{sec:fourier}
In order to most easily follow the evolution of the disk, it is useful to do a
Fourier transform on the azimuthal angle $\varphi$. We apply this to the mean
velocities and dispersion tensor elements after we calculated these in the usual
way (\ref{eq:moms}), except that this time we do not use a weighting function,
but take bins in $R$ and $\varphi$ to make sure every calculated trajectory
contributes to the final result. We use a discrete Fourier transform of the
following kind:
\begin{equation}
  \begin{array}{lll}
     c_m & = & \displaystyle
     \frac{2}{N} \sum_{j=1}^N f(\varphi_j) \cos m\varphi_j, \\[2ex]
     s_m & = & \displaystyle
     \frac{2}{N} \sum_{j=1}^N f(\varphi_j) \sin m\varphi_j,  
  \end{array}
\end{equation}
where $\varphi_j = j\pi/N$. This gives an approximate Fourier expansion
\begin{equation} \label{FourReconstruct}
  f(\varphi) \approx\frac{c_0}{2}+\frac{c_n}{2}\cos n\varphi
       +\sum_{m=1}^{n-1}(c_m \cos m\varphi + s_m \sin m\varphi),
\end{equation}
where $n=N/2$ and $N$ is supposed to be even. This has been done for $N=32$
in 25 radial bins.

We also construct an estimate of the surface density $\Sigma$ by dividing the
number of sampled orbits per bin by the segment area of the bin, and apply the
Fourier transform to this quantity as well.
\subsection{Error Estimation}
We employ a bootstrap method: the calculations leading from the data set of
every radial bin to its Fourier coefficients are redone for arbitrary subsamples
of this data set. The rms scatter in the outcome of many such calculations gives
a measure of the error.
\subsection{Symmetries and the Question of Stationarity}
In order to check, whether or how far our distributions have reached a
stationary equilibrium state, we employ a symmetry consideration. As noted by
\citet{fux01}, velocity distributions in an $m=2$ symmetric potential have a
symmetry
\begin{equation}
  \label{phisym}
  f(R, \varphi, u, v) = f(R, \pi - \varphi, -u, v),
\end{equation}
which is a consequence of the azimuthal $m=2$ symmetry and the time-reversal
symmetry of stellar dynamics in conjunction with stationarity of the
distribution function. We can turn this argument around and use the degree of
symmetry as a measure for stationarity. A quantitative measure of the symmetry
can be obtained in the obvious way by subtracting corresponding distribution
function values over a grid in phase space and doing a quadratic sum over the
grid points. Results of this kind of analysis are shown in Fig.\
\ref{fig:symtest}. As can be seen, the simulations show only a small increase in
asymmetry of 15\% above the noise level (as given by the initial state as well
as a very long-time simulation), and this is dropping rapidly after bar growth
is finished.
\section{Results} \label{sec:results}
\citet{kuij91} gave an analytical expression for the behaviour of the mean
velocities under the influence of a non-axisymmetric perturbation of multipole
order $m$ in a linear approximation. For our case, this yields
\begin{eqnarray} \label{uv_expect}
  \bar{u} & = & \frac{\alpha(R)\,v_0^2}{3R}
  \frac{3\Omega_b-\Omega}{(\Omega-\Omega_\ilr)(\Omega-\Omega_\olr)}
  \sin2\varphi,\nonumber \\ \\
  \bar{v} & = & v_0  - \frac{\alpha(R)\,v_0^2}{6R} 
  \frac{4\Omega_b-\Omega}{(\Omega-\Omega_\ilr)(\Omega-\Omega_\olr)} 
  \cos2\varphi, \nonumber 
\end{eqnarray}
where $\Omega_\ilr$ and $\Omega_\olr$ refer to the circular frequencies at the
radii of the inner and outer Lindblad resonance. For a flat rotation curve,
$\Omega=v_0/R$. In our range of interest, Eqs.~(\ref{uv_expect}) are dominated
by a pole at the OLR.

The results of our simulation, presented below, agree roughly with these
expectations: the mean velocities show modulations $\bar u\propto\sin2\varphi$,
$\bar v\propto\cos2\varphi$. The sign of the modulations varies with radius,
sign changes should indicate resonances. For the dispersion tensor the situation
is similar: the off-diagonal component shows a $\sin2\varphi$ perturbation, the
diagonal components a $\cos2\varphi$.
\subsection{Fourier Analysis}
Since we have an azimuthal $m=2$ symmetry, all coefficients of odd index $m$ are
expected to vanish, what they do within their errors. Furthermore, in the
undeveloped state before bar growth, all coefficients except those with $m=0$
are, by construction, zero within their errors.
\subsubsection{The $m=0$ Components:}
\begin{figure*}
  \centering \includegraphics[width=175mm]{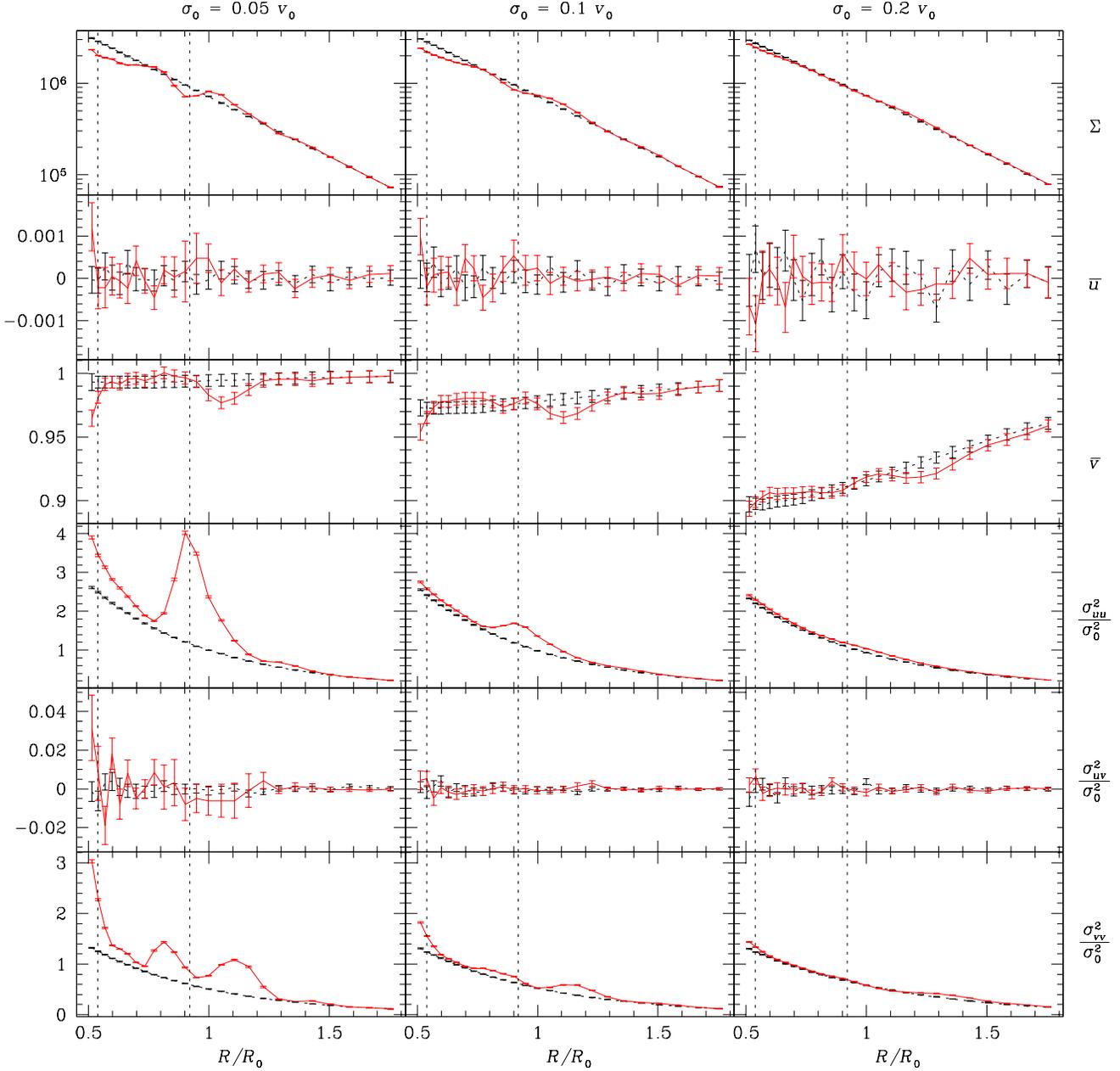}
  \caption{Axisymmetric ($m=0$) components of
    velocity distribution in the undeveloped state before bar growth ({\em
      dashed}) and in the final state ({\em solid}), plotted against radius.
    Results for models with initial $\sigma_0=0.05 v_0$, $0.1v_0 $ and $0.2v_0$
    are shown in the {\em left}, {\em middle} and {\em right} panels,
    respectively. Velocities are given in units of $v_0$ and $\Sigma$ in
    arbitrary units. The model shown has an OLR-radius of $0.92 R_0$,
    corresponding to a CR of $0.539 R_0$ ({\em dashed vertical lines}).}
\label{fig:zerocomps}
\end{figure*}
The dashed lines in Fig.\ \ref{fig:zerocomps} represent the initial axisymmetric
case, which shows an exponential decline of $\Sigma$ as well as in the diagonal
components of velocity dispersion tensor. The mean azimuthal motion deviates
from the circular speed $v_0$ by the asymmetric drift, which is stronger for
large $\sigma_0$, as expected. The mean radial motion vanishes, as required for
any stationary model.

The solid lines in Fig.\ \ref{fig:zerocomps} show the radial run of the $m=0$
components in the barred case. Apart from the components for $\bar u$ and
$\sigma^2_{uv}$, which have to vanish for stationary models, some signs of
perturbation are visible, the stronger the smaller $\sigma_0$.

First of all, we learn that dispersive effects are quite efficient in drawing
resonance-induced features away from the actual position of the resonance.
Whereas for small velocity dispersion the association of the features with the
OLR at $R_\olr=0.92 R_0$ is clearly visible, they appear in the high dispersion
case at a radial range of $1.1$ to $1.4 R_0$, where naively one would not
attribute them to the OLR.

In particular, we have a bump in the $\bar{v}$-curve outside of OLR,
which becomes more explicit with decreasing dispersion while roughly
keeping its absolute magnitude.  In contrast to this, the asymmetric
drift as the dominant deviation of $\bar{v}$ from the nominal rotation
velocity $v_0$ diminishes with decreasing dispersion.  There is a
similar bump in $\sigma_{uu}^2$, whereas we have a two-fold feature in
$\sigma_{vv}^2$.

Most of the perturbative features in the $m=0$ components are induced by the
opposite orientation of the near-circular orbits on either side of the
resonance.
\subsubsection{The $m=2$ Components}
\begin{figure*}
  \centering\includegraphics[width=175mm]{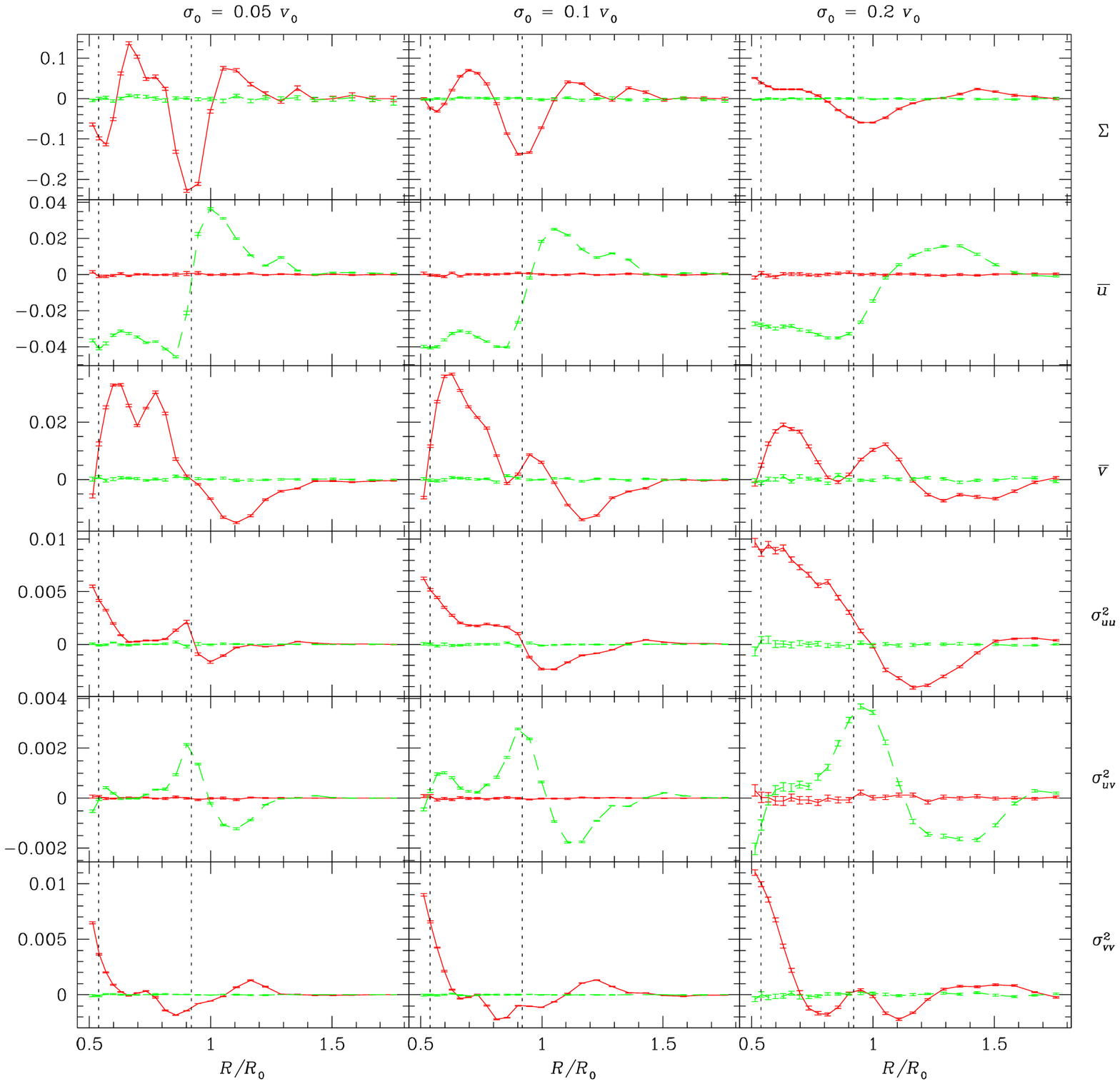}
  \caption{$m=2$ Fourier cosine ({\em solid})
    and sine ({\em dashed}) components, plotted against radius. Panels as in
    Fig.\ \ref{fig:zerocomps}. For surface density $\Sigma$, the value shown is
    relative to the undisturbed $m=0$ value. OLR and CR are at $0.92 R_0$ and
    $0.539 R_0$ respectively ({\em dashed vertical lines}).}
  \label{fig:twocossin}
\end{figure*}
\begin{figure}
  \centering \includegraphics[width=80mm]{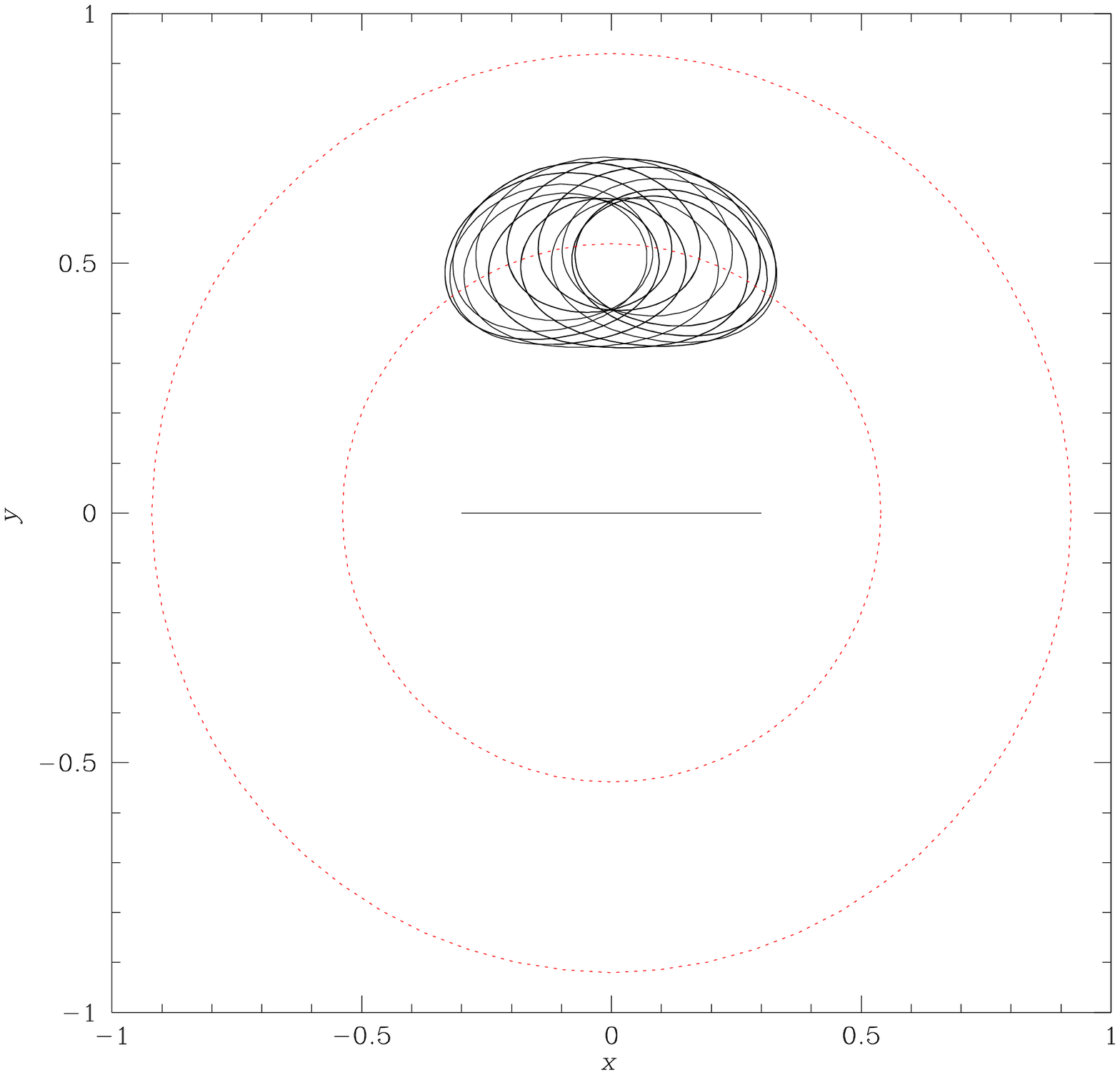}
  \caption[]{Orbit near co-rotation that circulates the L$_4$ point in
    a retrograde sense (anti-clockwise for a clockwise rotating bar).  The bar
    is indicated by a line, while the dotted circles correspond to CR and OLR.
    In our model, a considerable fraction of stars is trapped in orbits of this
    family, resulting in the somewhat peculiar run of the velocity moments
    (Fig.\ \ref{fig:twocossin}).}
  \label{fig:Lorbit}
\end{figure}

Because of the symmetry of the problem, we expect the $m=2$ components to be the
dominant ones.  As for every $m>0$, we have one more degree of freedom here,
since we have a cosine- and a sine-component (see Fig.\ \ref{fig:twocossin}), or
equivalently, an amplitude and a phase.

The cosine-terms of $\bar{u}$ and $\sigma^2_{uv}$ as well as the sine terms of
$\Sigma$, $\bar{v}$, $\sigma^2_{uu}$, and $\sigma^2_{vv}$ vanish (within their
error), which is to be expected from Eq.~(\ref{phisym}) for a stationary model.
The behaviour of the remaining non-vanishing terms agrees roughly with the
expectation from simple linear theory, cf.\ Eq.~(\ref{uv_expect}). However,
there are two significant deviations. First, as $\sigma_0$ increases (from left
to right panels in Fig.\ \ref{fig:twocossin}), the resonant features are shifted
away from $R_\olr$ to larger radii and are also somewhat smoothed out. Second,
there is an additional feature at $R\sim0.6R_0$, in particular, for $\bar{v}$.
When inspecting the orbits of stars dominating at this radius, we find that many
belong to an orbit family associated with the stable Lagrange points L$_4$ and
L$_5$, see Fig.\ \ref{fig:Lorbit}. These orbits can reach radii far beyond
co-rotation and, in a frame co-rotating with the bar, perform a retrograde
motion around the Lagrange points. They thus result in a reduced $\bar{v}$ at
azimuths perpendicular to the bar and hence lead to a positive $\cos2\varphi$
component.

\subsubsection{Higher-Order Components}
The $m=4$ components are usually smaller by at least a factor 3 compared to the
$m=2$ modes, but they are significantly different from zero. In contrast, the
analytical model (\ref{uv_expect}) had no excitation of higher modes at all,
since mode coupling is not included in a linear approximation.

The excitation of the $m=4$ modes follows the overall pattern seen in the $m=2$
case: in accordance with the symmetry requirements, we have a $\sin(4\varphi)$
in $u$ and $\sigma_{uv}^2$, and a $\cos(2\varphi)$ in all the other
components. Our  signal-to-noise ratio is in the range of about 3 for the $m=4$
case, and less for higher components, so surely the significance of our model
is petering out. Nevertheless, some $m=8$ modes do seem to be discernible still.
\subsection{Variation of Parameters}
Doubling the bar strength leads to a doubling of the magnitude of the effects.
This is true for the differences between the undeveloped and final state as well
as for Fourier coefficients.

Changing the bar size, i.e. the parameter $R_b$ in (\ref{barpot}), produces some
local variations in the magnitude of the effects, but it does not shift them in
location. Of course, this is to be expected if the location is determined by
resonance conditions.

Variation of the disk scale length does not seem to have any strong effects,
which is to be expected, since it only slightly affects the sampling of the
orbits but not the dynamics.

The shape of the velocity curve has a mild effect: we tried models with a
slightly rising ($\beta=0.1$) or falling ($\beta=-0.1$) velocity profile, and
found very similar results as for $\beta=0$. (Note that we kept the OLR radius
fixed, so models with different velocity profiles have different rotation
frequencies of the bar.) This is expected, too, since $\beta$ essentially
controls the distances between various resonances (growing wider for larger
$\beta$), while most of our results are dominated by a single resonance, the
OLR.
\subsection{The Axis Ratio of the Velocity Ellipsoid}\label{sec:axrat}
\begin{figure*}
  \centering \includegraphics[width=170mm]{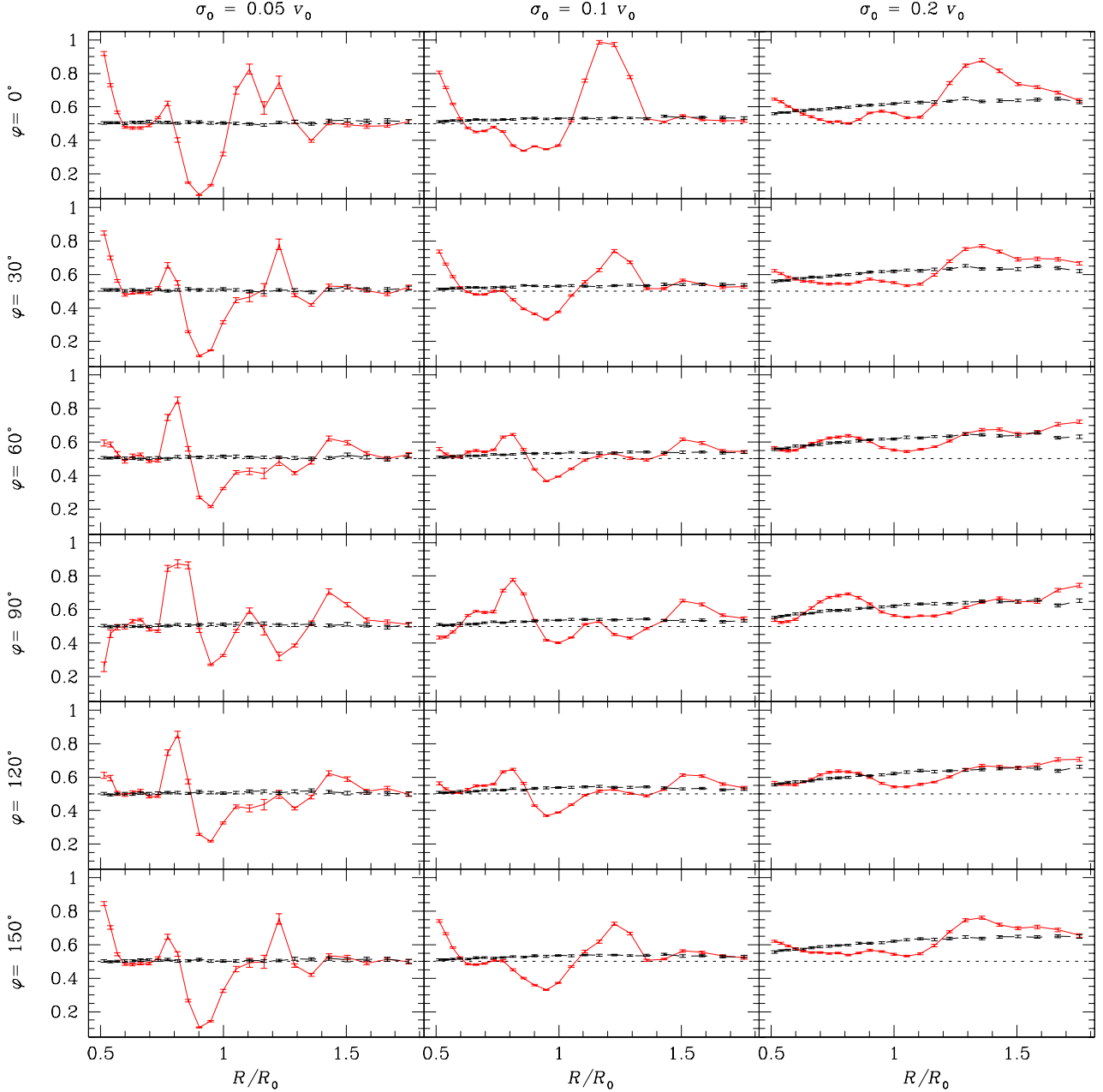}
  \caption{Axis ratio $\sigma^2_{2}/\sigma^2_{1}$ of the principal components of
    the velocity-dispersion tensor for the final (\emph{solid}) and the initial
    unperturbed state (\emph{dashed}) at various azimuths for the models with
    $R_\olr=0.92 R_0$ and $\sigma_0$ of $0.05v_0$ ({\em left}), $0.1v_0$
    (\emph{middle}), and $0.2v_0$ (\emph{right})
  \label{fig:axrat}}
\end{figure*}

Fig.\ \ref{fig:axrat} shows the ratio of the principal axes of the velocity
dispersion ellipsoid, i.e.\ the eigenvalues $\sigma^2_{1,2}$ of the tensor
$\boldsymbol{\sigma}^2$. For a flat rotation curve, this ratio is often expected
to be 0.5. Indeed, in the undisturbed case (for which $\sigma_1=\sigma_{uu}$ and
$\sigma_2=\sigma_{vv}$)
\begin{equation}
  \lim_{\sigma \rightarrow 0} \frac{\sigma_{2}^2}{\sigma_{1}^2} =
  \frac{\kappa^2}{4\Omega^2} \equiv 
  \frac{1}{2} \left( 1 + \frac{d \ln v_c}{d \ln R} \right),
\end{equation}
where $\kappa$ is the epicycle frequency. This relation is often given without
the limit and then referred to as Oort's relation. It is, however, important to
note that the error in Oort's relation is considerable already for $\sigma_0=0.2
v_0$ \citep{ec93,de99a}, which means that using it for the old stellar disk of
the Milky Way is, at best, dangerous. This can be seen clearly from Fig.\
\ref{fig:axrat} showing that $\sigma_{2}^2/\sigma_{1}^2$ for the unperturbed
case (dotted) significantly deviates from 0.5 for a warm stellar disk
($\sigma_0=0.2 v_0$).

After formation of the bar perturbation, we have large variations in this
quantity. For $\sigma_0=0.2v_0$ and at $R=R_0$, the values are generally
somewhat higher than Oort's value 0.5 but smaller than for an undisturbed
disk. For directions that are roughly along the bar ($\varphi\sim0$), this ratio
rises sharply just outside the solar circle to values reaching as high as 0.8.

It is instructive to take a look at the same quantities for a smaller velocity
dispersion of $\sigma_0=0.05v_0$ or $0.1v_0$ (left and middle panels of Fig.\
\ref{fig:axrat}). First of all, the axis ratio in the undisturbed case (dashed
line in the right panels) is much closer to Oort's value of 0.5 here, as
expected. Generally, the observed features have smaller width, i.e.\ they are
less washed out by dispersion, but the effect of the bar, i.e.\ the difference
between the solid and dashed lines, is still even quantitatively similar for the
different velocity dispersions. Thus, the deviation of $\sigma_{2}^2 /
\sigma_{1}^2$ from Oort's value may be decomposed into a velocity-dispersion
dependent term, which consists of a general elevation of only mild radial
dependence and a bar-induced term, which varies spatially and appears to be
negative for the Solar position.

As a consequence, for mildly warm stellar disks ($\sigma\la0.2v_0$), the ratio
$\sigma_{2}^2/\sigma_{1}^2$ in the Solar neighbourhood may well drop below
Oort's value.
\begin{figure*}
  \centering \includegraphics[width=170mm]{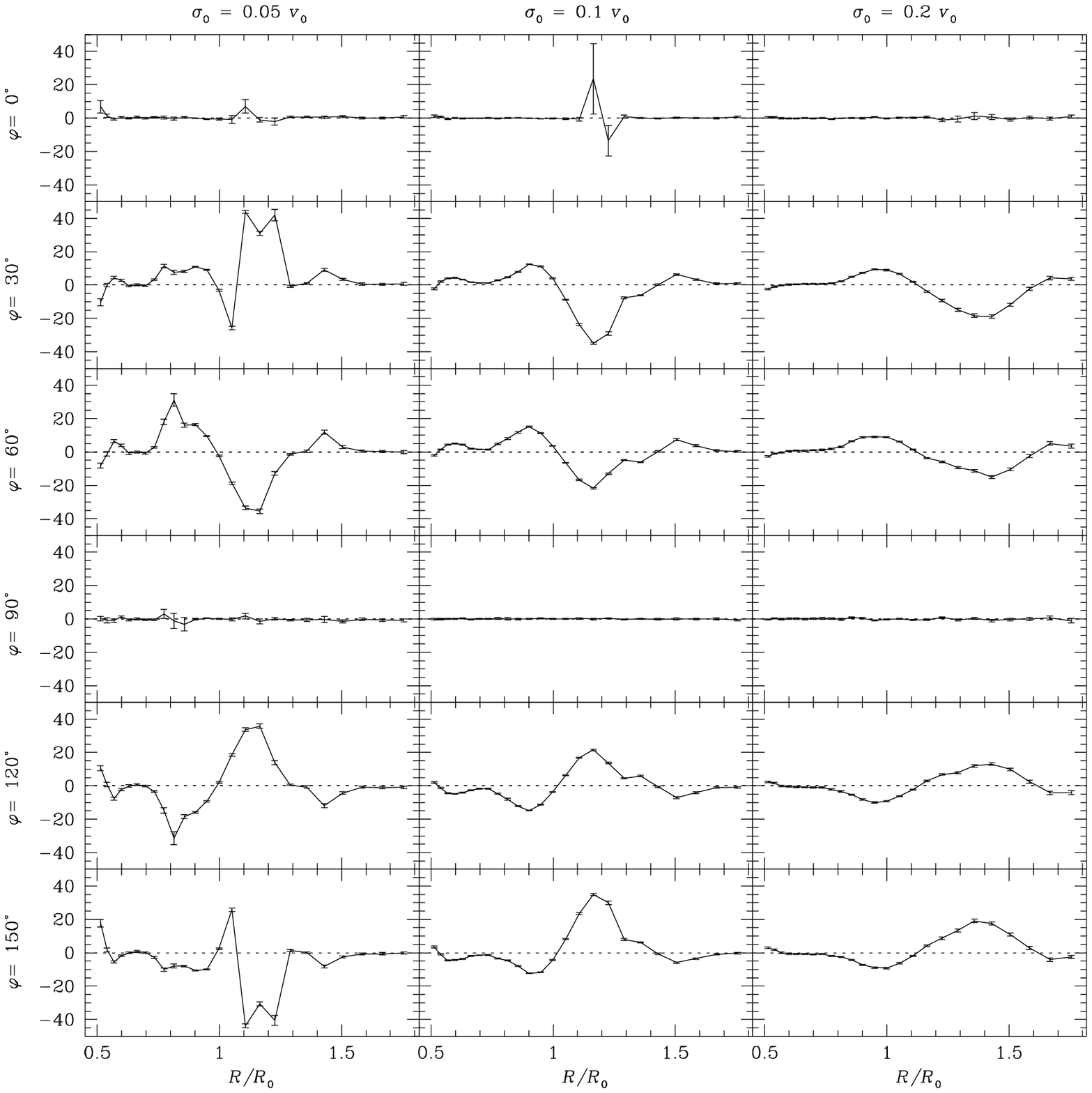}
  \caption{Vertex deviation $\ell_v$ at various azimuths for the same
    models as in Fig.\ \ref{fig:axrat}. The highly uncertain outliers at
    $R\sim1.2R_0$, $\varphi=0\degr$ for $\sigma_0=0.1v_0$ are caused by a
    near-circular velocity ellipsoid ($\sigma_1\sim\sigma_2$, see Fig.\
    \ref{fig:axrat}), which renders the vertex deviation ill-defined.}
  \label{fig:vertex}
\end{figure*}
\subsection{Vertex Deviation}
The so-called vertex deviation 
\begin{equation}
  \ell_v = \frac{1}{2}\arctan
  \frac{2\sigma_{uv}^2}{\sigma_{uu}^2-\sigma_{vv}^2}
\end{equation}
is the angle between the direction of the largest velocity dispersion and the
line to the Galactic center, see also Fig.\ \ref{fig:coorddef}. For
axisymmetric equilibrium models, $\ell_v=0$.

In Fig.\ \ref{fig:vertex}, we plot $\ell_v$, computed from Fourier coefficients
up to $m=2$ only in order to eliminate short-scale fluctuations. We find that
the bar-induced vertex deviation decreases with increasing velocity dispersion
from up to $\sim30\degr$ for $\sigma_0=0.05v_0$ to $5\degr$ -- $10\degr$ for
$\sigma_0=0.2v_0$. In the direction of the bar and perpendicular to it,
$\ell_v=0$, as expected from symmetry. For moderately warm stellar disks
($\sigma_0\ga0.1v_0$) the vertex deviation at azimuth angles between $0^\circ$
and $90^\circ$ (including the Solar azimuth) is positive at and inside the solar
circle and negative for some range outside. The actual radius where the sign
change occurs is obviously coupled to the OLR, but again shifted to the outside.
For a cold stellar disk, the bar-induced vertex deviation displays are more
complicated pattern and may reach as high as $40\degr$.

Obviously, the vertex deviation is antisymmetric with respect to
$\varphi\mapsto(\pi-\varphi)$, whereas the velocity dispersion axis ratio is
symmetric.  This is a simple consequence of symmetry (\ref{phisym}), which in
particular implies $(\sigma^2_{uu},\sigma^2_{uv},\sigma^2_{vv})
\mapsto(\sigma^2_{uu},-\sigma^2_{uv},\sigma^2_{vv})$.
\begin{figure*}
  \centering \includegraphics[width=\textwidth,clip]{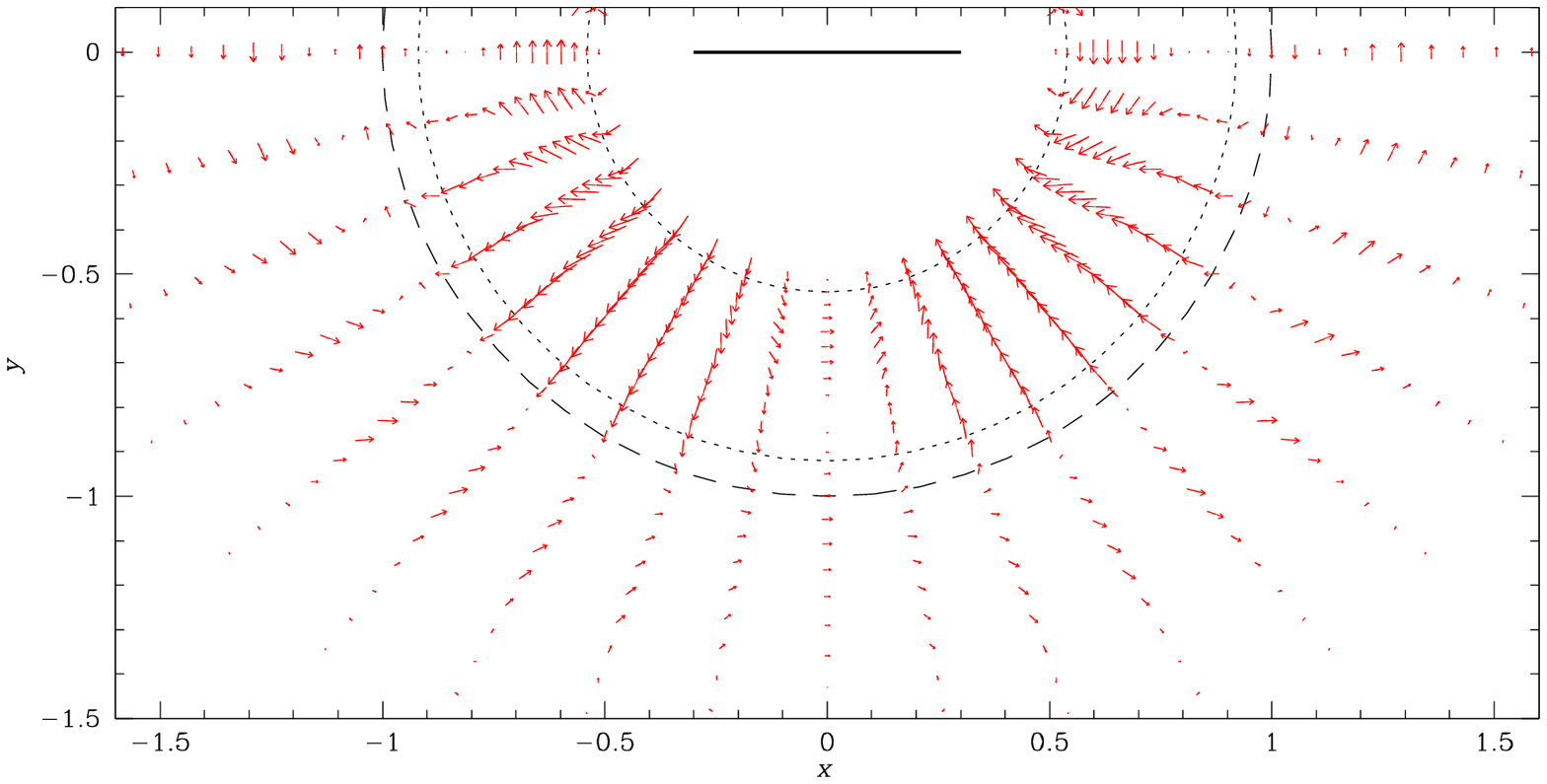}
  \caption{Bar-induced deviations of the mean velocity from the unperturbed
    state up to $m=2$ for $\sigma_0=0.2v_0$. The bar is aligned with the
    horizontal axis and is supposed to rotate clockwise. The solar circle is
    dashed, dotted circles are CR and OLR.}
  \label{fig:velfield}
\end{figure*}
\section{Discussion} \label{sec:disc}
We have studied the bar-induced variations of the mean velocities and velocity
dispersions in the outer disk of a barred galaxy with special emphasis on the
situation in the Milky Way. We generally find that these variations are largest
near the outer Lindblad resonance (OLR), the apparent radius of which, however,
may be shifted outwards by 10-20\% due to the non-circularity of the orbits in a
warm stellar disk.

Radial motions $\bar{u}$ of standards of rest can be seen to occur quite
frequently (see Fig.\ \ref{fig:velfield}), and can reach magnitudes of the
order of about $0.02 v_0$, corresponding to about 5 km\,s$^{-1}$ for the Milky
Way. Because of the $\sin2\varphi$ dependence, these would be maximal at
$\varphi=45\degr$, which is quite near the proposed position of the Sun of
$\varphi\approx30\degr$.  In its radial dependence however, $\bar{u}$ swings
through zero shortly outside of the OLR, and it may well be that the Sun just
meets that point. So we cannot give a definite prediction for the bar-induced
radial motion of the LSR here, not even by sign only, except that it should be
very small (at most a few km\,s$^{-1}$).

Variations of the mean azimuthal velocity are also present, but here we have to
deal with several perturbation effects, the dominant one for realistic cases
being traditional axisymmetric asymmetric drift. We can construct a picture of
the bar-induced $\bar{\B{v}}$-modifications only (see Fig.\ \ref{fig:velfield})
by subtracting $\bar{\B{v}}$ in the undeveloped case before bar growth, thus
also canceling asymmetric drift. However, this is, of course, impossible for
real galaxies, where only the combination of circular speed (LSR motion),
asymmetric drift, and bar induced drift is measurable (in principle). In
observational data, the effect of asymmetric drift can be corrected for, as done
by \citet{db98}, by calculating $\bar{v}$ for stellar populations with different
velocity dispersions and extrapolating to vanishing dispersion. This approach,
however, cannot be applied to the bar-induced perturbations in $\bar{v}$, since,
as can be seen in Fig.\ \ref{fig:zerocomps}, the relative magnitude of the
bar-induced wiggle keeps constant at about 2\%, independent of velocity
dispersion. So, as long as only local measurements are available, there is no
way to discern the bar-induced azimuthal velocity perturbations from variations
in the background rotation curve, and so it will be rather hard to draw any
observational constraints from the mean azimuthal velocities.

For an axisymmetric equilibrium, the no vertex deviation, i.e.\ $\ell_v=0$, is
expected, while in the Solar neighbourhood $\ell_v$ is observed to drop from
$\ell_v\approx30\degr$ for blue to $\approx 10\degr$ for red main-sequence stars
\citep{db98}. The traditional interpretation of these vertex deviations is that
young stars deviate from equilibrium, and thus even in an axisymmetric galaxy
give rise to $\ell_v\neq0$, which due to the decreasing number of young stars at
later stellar types is a decreasing function of stellar colour.

Our calculations show that for an equilibrium in a barred Milky Way, with bar
orientation, strength and pattern speed consistent with other data, a vertex
deviation of the size and direction as observed emerges naturally. Moreover, we
also found that for dynamically cooler sub-populations, i.e.\ bluer or younger
stars, the bar-induced vertex deviation increases in amplitude, very similar to
the observed values.

This gives strong support for the hypothesis that the vertex deviation observed
in the Solar neighbourhood is predominantly caused by deviations from
axisymmetry rather than from equilibrium. This explanation also naturally
accounts for the fact that $\ell_v$ for young stars has the same direction as
for old ones, which with the traditional explanation would be a chance
coincidence.

The axis ratio of the (principal components of the) velocity dispersion tensor,
$\sigma_2^2/\sigma_1^2$, is clearly affected by the central bar. In particular,
values less than 0.5, Oort's value for a flat rotation curve, are possible
\citep[Oort's value is a \emph{lower} limit for an axisymmetric galaxy, see
Sect.~\ref{sec:axrat} and][]{ec93,de99a}.  This nicely fits to the values
inferred from HIPPARCOS data \citep{db98}, which give
$\sigma_2^2/\sigma_1^2\sim0.42$ for the old stellar disk.

In the Milky Way, as well as in many barred galaxies, the bar is not the only
deviation from a smooth axisymmetric background. Most prominently, spiral arm
structure and an elliptic (or oval) disk and/or halo add further
non-axisymmetric perturbations. However, we will defer the discussion of the
influence of spiral arms on the outer disk kinematics to a future paper.

\begin{acknowledgements}
  We thank Marc Verheijen for a careful reading of the manuscript.
\end{acknowledgements}


\end{document}